\documentclass[11pt, letter]{article}
\usepackage{jheppub}
\usepackage[english]{babel}
\usepackage{blindtext}\blindmathtrue
\usepackage{avant} 
\usepackage[dvipsnames]{xcolor}
\usepackage{amsmath,epsf,amssymb,mathtools,latexsym,amsthm,setspace,array,pifont,hyperref,amsfonts,dsfont,cancel,braket,parskip,slashed}
\usepackage[none]{hyphenat} 
\usepackage{graphicx}
\usepackage{float}
\makeatletter
\graphicspath{{Pictures/}} 

\newcommand{\pd}{\partial}

\newcommand{\cl}[1]{\mathcal{#1}}

\def\prd{\ref@{Phys.~Rev.~D}}        

\usepackage{tcolorbox}
\definecolor{airforceblue}{rgb}{0.36, 0.54, 0.66}
\definecolor{azure}{rgb}{0.0, 0.5, 1.0}
\newtcolorbox{tdbox}{colback=airforceblue!40!white,colframe=azure!90!black} 
\newcommand{\td}[1]{
	\if\notesOn1
	\begin{tdbox}
		#1
	\end{tdbox}
	\fi
}
\hypersetup{
	colorlinks = true,
	linkcolor = Mahogany,
	anchorcolor = Mahogany,
	citecolor = Mahogany,
	filecolor = Mahogany,
	urlcolor = Mahogany
}

\def\notesOn{1}

\title{Massive Covariant Colour-Kinematics in 3D}
\author{Nathan Moynihan}
\affiliation[]{Higgs Centre for Theoretical Physics, School of Physics and Astronomy, \\The University of Edinburgh, EH9 3FD, Scotland}
\affiliation[]{School of Mathematics \& Hamilton Mathematics Institute,\\Trinity College Dublin, College Green, Dublin 2, Ireland}
\abstract{
We explore topologically massive gauge theories using the covariant colour kinematics duality recently introduced in \cite{Cheung:2021zvb}. We show that the massive bi-adjoint scalar field is simply related to topologically massive gauge theory by the duality, and that enacting the same duality on the gauge theory produces topologically massive gravity coupled to a scalar or, equivalently, an antisymmetric field. We also show that different choices for the replacement of the colour structure constants with kinematic structure constants lead to different theories, including a topologically massive generalisation of Born-Infeld theory. 
}

\usepackage{enumitem,amssymb}
\newlist{todolist}{itemize}{2}
\setlist[todolist]{label=$\square$}
\usepackage{pifont}
\begin{document}
\maketitle
\section{Introduction}
Colour-kinematics (CK) duality is a surprising relationship between seemingly distinct theories, relating gauge theories to gravity via an interesting web of double copies. This relationship was first understood in the context of scattering amplitudes \cite{Kawai:1985xq,Bern:2008qj,Bern:2010ue}, where the authors noted the remarkable fact that amplitudes in Yang-Mills theory, when formulated in a particular way, have kinematic numerators that obey the Jacobi relations satisfied by the amplitude's colour factors.  Furthermore, replacing said colour factors with copies of the kinematic factors immediately yields a gravitational amplitude, and further work has shown that colour algebras and kinematic algebras are related in certain cases \cite{Monteiro:2011pc,Bjerrum-Bohr:2012kaa,Monteiro:2013rya}. There are many examples of the basic double copy in action at the level of scattering amplitudes as well as classically in terms of the Kerr-Schild double copy \cite{Monteiro:2014cda,Luna:2015paa,Luna:2016due,Lee:2018gxc,CarrilloGonzalez:2019gof,Cho:2019ype,Carrillo-Gonzalez:2019aao,Bah:2019sda,Kim:2019jwm,Bahjat-Abbas:2020cyb,Luna:2020adi,Keeler:2020rcv,Elor:2020nqe,Gonzo:2021drq,Alkac:2021seh}, the Weyl double copy \cite{Luna:2018dpt,Alawadhi:2020jrv,Godazgar:2021iae,Chacon:2021wbr,Godazgar:2020zbv} and other interesting approaches \cite{Goldberger:2016iau,Goldberger:2017ogt,Adamo:2018mpq,Moynihan:2019bor,Casali:2020vuy,Adamo:2020qru,Shen:2018ebu,Easson:2020esh,Chacon:2020fmr,Cheung:2020djz,Cristofoli:2020hnk,Huang:2019cja,Moynihan:2020gxj,Adamo:2020syc,Alfonsi:2020lub,Alawadhi:2019urr,Bautista:2019evw,Banerjee:2019saj,Adamo:2021dfg,Chacon:2021hfe,Carrasco:2021bmu,Cheung:2017ems,Carrasco:2019qwr,Angus:2021zhy,Farnsworth:2021wvs,Borsten:2021hua,White:2020sfn,Casali:2020uvr,Emond:2020lwi,Ahmadiniaz:2021fey,Carrasco:2020ywq,Borsten:2020zgj}. The duality doesn't just relate gauge theories to gravity, however, and recent work on the duality has shown that there is a web of relationships between various theories, for example biadjoint scalars, the non-linear sigma model, Born-Infeld theory and special Galileon theory -- see \cite{Bern:2019prr} for a comprehensive review. Most progress in understanding colour-kinematics duality has been for massless theories. However, recent work has shown that CK duality applies to QCD with massive particles \cite{Johansson:2015oia,Johansson:2019dnu} and that CK duality is trivially observed up to four points for massive Yang-Mills in 4D \cite{Momeni:2020vvr,Johnson:2020pny} but fails at five points \cite{Johnson:2020pny} where the amplitudes become plagued with spurious poles.

It was recently conjectured that topologically massive gauge theories \cite{Deser:1981wh} satisfy colour-kinematics duality \cite{Moynihan:2020ejh}, and the basics of the double copy for some classical solutions was established \cite{Burger:2021wss}. It was also shown that tree-level topologically massive gluon and graviton scattering amplitudes do indeed exhibit colour-kinematic duality at 3, 4 and 5 points \cite{Gonzalez:2021bes}. In the matter-coupled case, it was discovered that the double copy of topologically massive gauge theory gives rise to topologically massive gravity (TMG) plus an extra propagating scalar mode \cite{Burger:2021wss}. This additional mode turned out to be crucial in obtaining the massless limit of the double copy, matching the classical solution previously derived in \cite{CarrilloGonzalez:2019gof}. It was found that the propagator of the double copied theory matches topologically massive gravity coupled to a scalar (either conformally \cite{Deser:1990bj} or as in string theory via the usual dilaton coupling \cite{Burger:2021wss}). This is perhaps not entirely surprising, since it is well known that the double copy can in general produce a graviton, a dilaton and a Kalb-Ramond field. To further explore the precise content of the topologically massive double copy, in this paper we will look at colour-kinematics from the perspective of the equations of motion, using the \textit{covariant colour-kinematics duality} recently introduced by Cheung and Mangan \cite{Cheung:2021zvb}.
\section{Massive Covariant Colour-Kinematics Duality}
In this section we will review the basics of the covariant colour-kinematics duality between the massless bi-adjoint scalar field and Yang-Mills, and we refer the reader to Ref \cite{Cheung:2021zvb} for further details. In order to study the duality, it is convenient to study the equations of motion in so-called second order form, as was pointed out in \cite{Cheung:2021zvb}. This is due to the fact that the Klein-Gordon equation for the bi-adjoint scalar field is second order in the covariant derivative while the gauge theory and gravity equations of motion are typically first order. For example in the source free case
\begin{equation}\label{key}
	D^2\phi^{a\bar{a}} = 0,~~~~~D_\mu F^{a\mu\nu} = 0,~~~~~\nabla_\mu R^{\mu\nu\rho\sigma} = 0.
\end{equation}
In order to enact the double copy at this level, it was suggested in \cite{Cheung:2021zvb} that the equations of motion should be reformulated into second order form. The authors then showed that there is a natural duality between colour and kinematics, which we will now review.
\subsection{Bi-adjoint Scalars and Yang-Mills}
We begin by defining the Lagrangian for a biadjoint scalar field and a source $J^{a\bar{a}}$
\begin{equation}\label{key}
	\mathcal{L}^{\mathrm{BAS}}=\frac{1}{2} \partial_{\mu} \phi^{a \bar{a}} \partial^{\mu} \phi^{a \bar{a}}-\frac{1}{3 !} f^{a b c} f^{\bar{a} \bar{b} \bar{c}} \phi^{a \bar{a}} \phi^{b \bar{b}} \phi^{c \bar{c}}+\phi^{a \bar{a}} J^{a \bar{a}} + \frac12 m^2\phi^{a\bar{a}}\phi^{a\bar{a}},
\end{equation}
where the structure constants $f^{abc}$ and generator of the colour algebra $T^a$ obey
\begin{equation}\label{colouralgebra}
	\left[T^{a}, T^{b}\right]=i f^{a b c} T^{c} \quad \text { and } \quad \operatorname{tr}\left[T^{a} T^{b}\right]=\delta^{a b}.
\end{equation}
This can be gauged by minimally coupling a gauge field in the `unbarred' sector, giving a Lagrangian for \textit{gauged} biadjoint scalar theory
\begin{equation}\label{key}
	\mathcal{L}^{\mathrm{GBAS}}=\frac{1}{2} D_{\mu} \phi^{a \bar{a}} D^{\mu} \phi^{a \bar{a}}-\frac{1}{3 !} f^{a b c} f^{\bar{a} \bar{b} \bar{c}} \phi^{a \bar{a}} \phi^{b \bar{b}} \phi^{c \bar{c}}+\phi^{a \bar{a}} J^{a \bar{a}} + \frac12 m^2\phi^{a\bar{a}}\phi^{a\bar{a}},
\end{equation}
where $D_\mu \phi^{a\bar{a}} = \pd_\mu \phi^{a\bar{a}} + f^{abc}A_\mu^b\phi^{c\bar{a}}$ and the dual colour is not gauged. The equation of motion is then
\begin{equation}\label{key}
	D^{2} \phi^{a \bar{a}}+\frac{1}{2} f^{a b c} f^{\bar{a} \bar{b} \bar{c}} \phi^{b \bar{b}} \phi^{c\bar{c}} + m^2\phi^{a\bar{a}} =J^{a \bar{a}}.
\end{equation}
This equation of motion is second order in the covariant derivative, so in order to enact the duality we will render the Yang-Mills equations of motion into a similar form. In first order form, they are given by
\begin{equation}\label{key}
	D^\mu F^{a}_{\mu\nu} = J^{a}_\nu,
\end{equation}
where the covariant derivative acts as $D_{\rho} F_{\mu \nu}^{a}=\partial_{\rho} F_{\mu \nu}^{a}+f^{a b c} A_{\rho}^{b} F_{\mu \nu}^{c}$.

Putting this into a form that's useful for the covariant double copy is a matter of acting with another covariant derivative $D_\rho$ and anti-symmetrizing to find
\begin{equation}\label{key}
	(D_\tau D^\mu F^{a}_{\mu\nu} - D_\nu D^\mu F^{a}_{\mu\tau}) = D_{[\tau}J^{a}_{\nu]},
\end{equation}
The first term is given by
\begin{align}\label{key}
	D_\tau D^\mu F^{a}_{\mu\nu} - D_\nu D^\mu F^{a}_{\mu\tau} &= D^\mu(D_\tau F^{a}_{\mu\nu} - D_\nu F^{a}_{\mu\tau}) + f^{abc}F_{\rho[\mu}^bF_{\nu]}^{c\rho}\\ 
	&= D^2F_{\tau\nu} + f^{abc}F_{\rho[\mu}^bF_{\nu]}^{c\rho},
\end{align}
where on the first line we have used $D_{[\rho} D_{\sigma]} F_{\mu \nu}^{a}=f^{a b c} F_{\rho \sigma}^{b} F_{\mu \nu}^{c}$ and on the second we have used the Bianchi identity. The equations of motion in this new second order form are then given by
\begin{equation}\label{key}
	D^2F_{\tau\nu}^a + f^{abc}F_{\rho[\tau}^bF_{\nu]}^{c\rho} = D_{[\tau}J^{a}_{\nu]} = J^a_{\tau\nu}.
\end{equation}

To enact colour-kinematics duality, we replace colour indices with kinematic spacetime indices, and $SU(N)$ structure constants with kinematic structure constants. If we replace a colour index with a single spacetime index, we find the second-order form of the NLSM \cite{Cheung:2021zvb}, where we identify the $f^{\mu\nu\rho}$ as the structure constant for the algebra of diffeomorphisms. However, as we are interested in the double copy to Yang-Mills, we replace each colour index with \textit{two} spacetime indices to find
\begin{equation}\label{key}
	D^{2} \phi^{a \bar{a}}+\frac{1}{2} f^{a b c} f^{\bar{a} \bar{b} \bar{c}} \phi^{b \bar{b}} \phi^{c\bar{c}}=J^{a \bar{a}} ~~~\longrightarrow~~~ D^{2} F^{a \mu\nu}+\frac{1}{2} f^{a b c} f^{\mu\nu\rho\sigma\tau\chi} F^{b}_{\rho\sigma}F^{c}_{\tau\chi} =D^{[\mu}J^{a \nu]}.
\end{equation}
We now need to interpret the object $f^{\mu\nu\rho\sigma\tau\chi}$, which we recognise as the structure constant associated to the algebra of the Lorentz group, i.e.
\begin{align}\label{spinlorentz}
	[S_{\mu\nu},S_{\rho\sigma}] &= 2f_{\mu\nu\rho\sigma\tau\chi}S^{\tau\chi} \\&= \left[\eta_{\mu\tau}\left(\eta_{\rho\chi}\eta_{\nu\sigma} - \eta_{\sigma\chi}\eta_{\nu\rho}\right) + \eta_{\nu\tau}\left(\eta_{\sigma\chi}\eta_{\mu\rho} - \eta_{\rho\chi}\eta_{\mu\sigma}\right)\right]S^{\tau\chi}.
\end{align}
Plugging this in, we find
\begin{equation}\label{key}
	D^{2} \phi^{a \bar{a}}+\frac{1}{2} f^{a b c} f^{\bar{a} \bar{b} \bar{c}} \phi^{b \bar{b}} \phi^{c\bar{c}}=J^{a \bar{a}} ~~~\longrightarrow~~~ D^{2} F^{a \mu\nu}+f^{a b c} F^{b\rho[\mu}F^{c\nu]}_{\rho} =D^{[\mu}J^{a \nu]},
\end{equation}  
which is indeed the second-order Yang-Mills EOM we found earlier.

The second order equations are equations of motion for the gauge-invariant field strength, rather than the underlying gauge field that we are normally interested in scattering. In standard quantum field theory, it is enough to know the one-point function in the presence of a source (typically derived from the generating functional $W[J]$) in order to derive $n$-point correlation functions. Such one-point functions are solutions to the equations of motion, and defined as
\begin{equation}\label{key}
	\left\langle\phi^{a \bar{a}}(p)\right\rangle_{J}=\frac{1}{i} \frac{\delta W[J]}{\delta J^{a \bar{a}}(p)}.
\end{equation}

Scattering amplitudes can then be derived from these correlation functions using the LSZ reduction prescription
\begin{equation}\label{LSZ}
	\left(\frac{i}{\sqrt{Z}}\right)^{n} \int \prod_{i=1}^{n} d x_{i}\left[e^{i (x_{i} \cdot q_{i})}\left(\pd^2_{x_{i}}+m^{2}\right)\left\langle\phi\left(x_{1}\right) \phi\left(x_{2}\right) \cdots \phi\left(x_{n}\right)\right\rangle_{\text {conn. }}\right].
\end{equation}
This tells us that if we can relate the one-point function of the field strength to the one-point function of the gauge field, we know that their scattering amplitudes must be related. One issue in doing this is that the field strength is gauge-invariant whereas the gauge field is not, and we are required to fix a gauge in order to relate the two quantities. Introducing an axial gauge vector $n^\mu$ (which enters the Lagrangian as $\sim (n_\mu A^{a\mu})^2$) with $n\cdot A^a \simeq \cl{O}(A^2)$, such that we can express the gauge field \textit{on-shell} as
\begin{equation}\label{gfixed}
	A_\mu^a = -\frac{n^\nu F_{\mu\nu}^a}{n\cdot\pd} = \tilde{\epsilon}^\nu F_{\mu\nu}.
\end{equation}
We can then relate the one-point functions as
\begin{equation}\label{key}
	\braket{A_\mu^a(q)}_{J} = \tilde{\epsilon}^\nu(q)\braket{F_{\mu\nu}^a(q)}_{J},
\end{equation}
and therefore the $n$-point functions by
\begin{equation}\label{key}
	\braket{A_{\mu_1}^{a_1}(q_1)A_{\mu_2}^{a_2}(q_2)\cdots A_{\mu_n}^{a_n}(q_n)}\bigg|_{J=0} = \left(\prod_{i=1}^{n-1} \frac{1}{i} \frac{\delta}{\delta J^{a_{i} \mu_{i}}\left(p_{i}\right)}\right)\tilde{\epsilon}^{\nu_n}(q_n)\braket{F_{\mu_n\nu_n}^{a_n}(q_n)}\bigg|_{J = 0},
\end{equation}
where both sides of this equation are taken to be on-shell and are gauge-invariant. This relationship ensures that scattering amplitudes in both theories are related. 
\subsection{Topologically Massive Yang-Mills}
Topologically massive Yang-Mills is described by the Lagrangian
\begin{equation}
	\cl{L} = -\frac{1}{4} F^a_{\mu\nu}F^{a\mu\nu}+\frac{m}{2}\, \epsilon^{\mu\nu\rho} \left(A^a_\mu \pd_\nu A^a_\rho+ \frac{1}{3} f^{abc} A^a_\mu A^b_\nu A^c_\rho\right).
\end{equation}
Varying this action with respect to $A_\mu^a$ gives the equations of motion
\begin{equation}\label{tymeom}
	D^\mu F^{a}_{\mu\nu} + \frac{m}{2}\epsilon_{\nu\rho\sigma}F^{a\rho\sigma} = J^{a}_\nu,
\end{equation}
where the covariant derivative acts as $D_{\rho} F_{\mu \nu}^{a}=\partial_{\rho} F_{\mu \nu}^{a}+f^{a b c} A_{\rho}^{b} F_{\mu \nu}^{c}$. The mass term which enters here is topological in the sense that it does not depend on the metric, being dependent only on the global topological properties of the underlying manifold. We will identify any such terms that arise in this paper as being topological. 

We suppress coupling constants throughout this paper, however they can simply be restored by dimensional analysis. Scattering amplitudes with $n$ external legs have dimension $\frac{n}{2}(2-D) + D$, and so amplitudes with an odd number of legs have a fractional mass dimension in $D=3$. Three-particle amplitudes, which we will examine shortly, will typically have couplings of the form $\tilde{g} = gm^\alpha$ where $[g] = \beta/2$ and $\alpha$ and $\beta$ are integers.

By performing exactly the same steps as we did in the last section, we can derive equations of motion of the form
\begin{equation}\label{covYMeom}
	D^2F_{\tau\nu}^a + f^{abc}F_{\rho[\tau}^bF_{\nu]}^{c\rho} + \frac{m}{2}D_{[\tau}\epsilon_{\nu]\rho\sigma}F^{a\rho\sigma} = D_{[\tau}J^{a}_{\nu]},
\end{equation}
which we recognise as the same second-order Yang-Mills equations with the addition of a topological mass term. 

It is not obvious by inspection how we could get such a topological mass term from the double copy of the bi-adjoint scalar with a standard mass term. However, in $2+1$ dimensions we can simplify things greatly by exploiting the fact that antisymmetric two-tensors are dual to vectors via Hodge duality. We can therefore define a dual field strength
\begin{equation}\label{key}
	F^{a\mu} = \frac12\epsilon^{\mu\nu\rho}F_{\nu\rho}^a.
\end{equation}
This dual vector, together with the relation
\begin{equation}\label{key}
	F^{a\mu}F^{b\nu} = -F^{a\mu}_\rho F^{b\rho\nu} + \frac12\eta^{\mu\nu}F_{\rho\sigma}^aF^{b\rho\sigma},
\end{equation}
and the Bianchi identity $D_\mu F^{a\mu} = 0$, leads to a new form of the equations of motion
\begin{equation}\label{key}
	D^{[\mu} F^{a\nu]} + m\epsilon^{\mu\nu\rho}F^{a}_\rho = \epsilon^{\mu\nu\rho}J^{a}_\rho.
\end{equation}
To derive a second order form of this equation, we can act on eq. \eqref{covYMeom} with $\epsilon^{\mu\tau\nu}$ and use the above three equations to find
\begin{equation}\label{TYMEOM}
	(D^2 + m^2)F^{\mu a}- f^{abc}\epsilon^{\mu\nu\rho}F_\nu^b F_{\rho}^c = \tilde{J}^{\mu a},
\end{equation}
where $\tilde{J}^{\mu a} =  \left(\epsilon^{\mu\nu\rho}D_\nu  + m\eta^{\mu\rho}\right)J_\rho^a$. 

In this form, these equations look suspiciously similar to the biadjoint scalar equations, but with some colour indices replaced with spacetime indices. Indeed, if we consider the most naive double copy possible, i.e. $\left\{\bar{a},\bar{b},\bar{c}\right\} \rightarrow \left\{\mu,\nu,\rho\right\}$, then we find
\begin{equation}\label{tymdc}
	D^{2} \phi^{a \bar{a}}+\frac{1}{2} f^{a b c} f^{\bar{a} \bar{b} \bar{c}} \phi^{b \bar{b}} \phi^{c\bar{c}} + m^2\phi^{a\bar{a}}=J^{a \bar{a}} ~~~\longrightarrow~~~ D^{2} F^{a\mu}+f^{a b c} f^{\mu\nu\rho} F^{b}_\nu F^c_\rho  + m^2F^{\mu a} =J^{a\mu}.
\end{equation}
We need to identify $f^{\mu\nu\rho}$ with a kinematic structure constant. In four dimensions, the kinematic structure constant was related to the spin-algebra of the Lorentz group, and this is equally true in 2+1 dimensions, since the spin-algebra in $D = 3$ is given by \cite{Jackiw:1990ka}
\begin{equation}\label{key}
	[J_\mu,J_\nu] = -\epsilon_{\mu\nu\rho}J^\rho.
\end{equation}
%

We have shown that the equations of motion of a biadjoint scalar and topologically massive Yang-Mills are simply related by colour-kinematic duality, at least in second order form. However, as we are interested in the scattering amplitudes, we need to ensure that the one-point functions can be related. Working in the axial gauge with gauge vector $n^\mu$ such that $n^\mu A_\mu^a = 0$, we can express $A^{a\mu}$ in terms of $F^{a\mu}$ via
\begin{equation}\label{key}
	A^{\mu a} = \frac{\epsilon^{\mu\nu\rho}n_\nu F_\rho^a}{n\cdot\pd} = \tilde{\epsilon}^{\mu\rho}F_{\rho}^a.
\end{equation}
This relationship is entirely unsurprising, since it is simply the result of dualising the field strength in eq. \eqref{gfixed}. We find then that the on-shell one-point functions are related via
\begin{equation}\label{key}
	\braket{A^{\mu a}(q)}_{J} = \tilde{\epsilon}^{\mu\nu}(q)\braket{F_{\nu}^a(q)}_{J},
\end{equation}
and therefore the $n$-point functions by
\begin{equation}\label{key}
	\braket{A_{\mu_1}^{a_1}(q_1)A_{\mu_2}^{a_2}(q_2)\cdots A_{\mu_n}^{a_n}(q_n)}\bigg|_{J=0} = \left(\prod_{i=1}^{n-1} \frac{1}{i} \frac{\delta}{\delta J^{a_{i} \mu_{i}}\left(p_{i}\right)}\right)\tilde{\epsilon}^{\mu_n\nu_n}(q_n)\braket{F_{\nu_n}^{a_n}(q_n)}\bigg|_{J = 0},
\end{equation}
which ensures that their scattering amplitudes are related. This condition is the 2+1 dimensional equivalent of the Yang-Mills condition we imposed in section 1 for massless Yang-Mills, i.e. the Hodge dual of eq. \eqref{gfixed}. We can, however, make an even simpler choice, since we can use the linearized equations of motion to relate the polarization of $A_\mu^a$ and $F_\mu^a$ in a different way, at least on-shell. At the linearized level, the topologically massive Yang-Mills equations of motion are
\begin{equation}\label{key}
	\pd^2 A_\mu^a + \pd\cdot A^a \pd_\mu + \frac{m}{2}F_\mu^a = 0.
\end{equation}
In the gauge $\pd\cdot A^a = 0$, we can relate one-point functions via
\begin{equation}\label{key}
	\braket{A^{\mu a}(q)}_{J} = \frac{1}{2m}\braket{F^{\mu a}(q)}_{J}.
\end{equation}

To illustrate the utility of this relationship, it is instructive to look at the three-particle on-shell amplitude. By inspection of the EOM in eq. \eqref{TYMEOM}, the three-particle vertex is found to be
\begin{equation}\label{key}
	V^{abc,\mu\nu\rho} = igmf^{abc}\epsilon^{\mu\nu\rho},
\end{equation}
where we have restored the coupling by dimensional analysis.
In momentum space, we can take the asymptotic states of the fields to be plane waves of the form
\begin{equation}\label{key}
	A_\mu^a(q) = c^a\epsilon_\mu(q)e^{iq\cdot x},~~~~~F_\mu^a(q) = 2mc^a\tilde{\epsilon}_\mu(q)e^{iq\cdot x}.
\end{equation}
We find then that the three-particle amplitude is given by
\begin{equation}\label{key}
	\cl{A}_3^{F^3} = 6igm^4f^{abc}\varepsilon^{\mu\nu\rho}\epsilon^1_\mu\epsilon_\nu^2\epsilon_\rho^3 = 6m^3\cl{A}_3^{A^3}.
\end{equation} 
$\cl{A}_3^{A^3}$ is precisely the three-particle amplitude arrived at via little group scaling and spinor-helicity methods \cite{Moynihan:2020ejh} or after much simplification of the traditional Feynman rules \cite{Gonzalez:2021bes}. Note that we could have proceeded as in \cite{Cheung:2021zvb} and introduced the notion of root and leaf legs, summed over the configurations and then utilised the various 3D simplifications to arrive at the same result. However, the straightforward kinematics in 2+1 dimensions meant that this was unnecessary since the one-point functions between field strength and gauge field are trivially related by Hodge duality and a rescaling. It should also be apparent that the Feynman rules are actually \textit{simpler} in the second-order form than in first-order form, and so deriving the scattering amplitudes should in general be much easier.

\subsection{Topologically Massive Gravity}
Let's now play the same game with gravity in the hope that we will be rewarded with similar success. Topologically massive gravity is described by the action
\begin{equation}\label{key}
	S = \frac{1}{2 \kappa^{2}} \int d^{3} x \sqrt{-g}\left[R+\frac{1}{2 m} \epsilon^{\lambda \mu \nu} \Gamma_{\lambda \sigma}^{\rho}\left(\partial_{\mu} \Gamma_{\nu \rho}^{\sigma}+\frac{2}{3} \Gamma_{\mu \tau}^{\sigma} \Gamma_{\nu \rho}^{\tau}\right)\right],
\end{equation}
and the equations of motion are given by
\begin{equation}\label{tmgeom}
	G_{\mu\nu} + \frac{1}{m}C_{\mu\nu} = T_{\mu\nu},
\end{equation}
where $C^{\mu\nu} = \frac{1}{\sqrt{g}}\epsilon^{\mu\alpha\rho}D_\alpha S^\nu_\rho$ is the Cotton-York tensor with $S^\nu_\rho$ the Schouten tensor defined by
\begin{equation}\label{key}
	S_{\mu \nu}=\frac{1}{D-2}\left(R_{\mu \nu}-\frac{1}{2(D-1)} g_{\mu \nu} R\right).
\end{equation}
We can express this in first-order form in terms of a covariant differential operator which depends only linearly on the covariant derivative, yielding
\begin{equation}\label{key}
	\cl{O}_{\mu\nu}^{~~~\rho\sigma}(m)R_{\rho\sigma} = T_{\mu\nu},
\end{equation}
where the operator $\cl{O}_{\mu\nu}^{~~~\rho\sigma}(m)$ is given by
\begin{equation}\label{key}
	\cl{O}_{\mu\nu}^{~~~\rho\sigma}(m) = \left(\delta_{\mu}^{\rho} \delta_{v}^{\sigma}-\frac{1}{2} g_{\mu \nu} g^{\rho \sigma}\right)+\frac{1}{m \sqrt{g}} \varepsilon_{\mu}{ }^{\alpha \beta}\left(\delta_{\beta}^{\lambda} \delta_{\nu}^{\sigma}-\frac{1}{4} g^{\rho \sigma} g_{\nu \beta}\right)D_\alpha.
\end{equation}
To derive the equations of motion in second order form, we apply this operator again with $m\rightarrow -m$ to find
\begin{equation}\label{key}
	\cl{O}_{\alpha\beta}^{~~~\mu\nu}(-m)\cl{O}_{\mu\nu}^{~~~\rho\sigma}(m)R_{\rho\sigma} = \cl{O}_{\alpha\beta}^{~~~\mu\nu}(-m)T_{\mu\nu}.
\end{equation}
Using the fact that the trace of eq. \eqref{tmgeom} gives $R = -2T$ and working through the algebra, we eventually find
\begin{equation}\label{key}
	(D^2+m^2)R_{\mu\nu} + g_{\mu\nu}R^{\rho\sigma}R_{\rho\sigma} - 3R^{~\rho}_{\mu}R_{\nu\rho} = J_{\mu\nu},
\end{equation}
where $J_{\mu\nu}$ is some complicated function of the stress-energy tensor and its trace\footnote{See eq. 3.28 in \cite{Burger:2021wss} for the full expression. For a traceless source, this is just the operator acting on the stress-energy tensor \begin{equation}\label{key}
		J_{\mu\nu} = \cl{O}_{\alpha\beta}^{~~~\mu\nu}(-m)T_{\mu\nu}.
\end{equation}}, and we used the fact that the Riemann tensor can be decomposed (in any dimension) as
\begin{equation}\label{riemanndecomp}
		R_{\mu \nu \alpha \beta}=W_{\mu \nu \alpha \beta}+\left(g_{\mu \alpha} S_{\nu \beta}-g_{\mu \beta} S_{\nu \alpha}-g_{\nu \alpha} S_{\mu \beta}+g_{\nu \beta} S_{\mu \alpha}\right),
\end{equation}
and in $D = 3$ we have $W_{\mu \nu \alpha \beta} = 0$.
It is instructive to look at the free-field equations at this point. Linearizing the second-order EOM around flat space in the harmonic gauge\footnote{In the harmonic gauge, the linearized Ricci tensor is given by $$R_{\mu\nu} = -\frac12\pd^2 h_{\mu\nu}.$$}, we find the expected equations of motion at linear order \cite{Deser:1981wh,Burger:2021wss}
\begin{equation}\label{key}
	(\pd^2 + m^2)\pd^2 h_{\mu\nu} = 0.
\end{equation}

We have derived two forms of the equations of motion for topologically massive gravity, and we will examine both. We will look at the first order form to begin with, where we make the colour index replacements on the eom \eqref{tymeom} minimally coupled to gravity, i.e.
\begin{equation}
	\nabla^\mu F^{a}_{\mu\nu} + f^{abc}A^{b\mu} F^c_{\mu\nu} +  \frac{m}{2}\epsilon_{\nu\rho\sigma}F^{a\rho\sigma} = J^{a}_\nu,
\end{equation}
where $\nabla^\mu$ is the usual gravitational covariant derivative. Replacing the colour indices $a \rightarrow \alpha\beta$ gives 
\begin{equation}
	{\nabla'}^\mu R_{\mu\nu\alpha\beta} + f^{\alpha\beta\gamma\tau\chi\omega}\omega_{\gamma\tau}^{\mu} R_{\chi\omega\mu\nu} +  \frac{m}{2}\epsilon_{\nu\rho\sigma}R^{\rho\sigma}_{~~\alpha\beta} = \nabla_{[\alpha} T_{\beta]\nu} + m\epsilon_{\alpha\beta}^{~~\mu} T_{\mu\nu},
\end{equation}
where we note that $\nabla'^\mu$ is the covariant derivative acting only on the first two indices of the Riemann tensor. We can identify $f^{\alpha\beta\gamma\tau\chi\omega}$ with the Lorentz algebra in eq. \eqref{spinlorentz}, and we find that this gives a simpler form in terms of the full gravitational covariant derivative
\begin{equation}
	{\nabla}^\mu R_{\mu\nu\alpha\beta} +  \frac{m}{2}\epsilon_{\nu\rho\sigma}R^{\rho\sigma}_{~~\alpha\beta} = \nabla_{[\alpha} J_{\beta]\nu} + m\epsilon_{\alpha\beta}^\mu T_{\mu\nu}.
\end{equation}

To see that this is an equation of motion for topologically massive gravity, we can act on it with $-\frac{1}{2m}\epsilon^{~\alpha\beta}_\lambda$ to find
\begin{equation}\label{dceom}
	G_{\mu\nu} + \frac{1}{m}C_{\mu\nu} + \frac{1}{m}\epsilon_{\mu\nu\rho}\nabla_\lambda S^{\lambda\rho} = \frac{1}{m}\epsilon^{~\alpha\beta}_\mu \nabla_{\alpha} J_{\beta\nu} + T_{\mu\nu},
\end{equation}
where we have used the Bianchi identity
\begin{equation}\label{key}
	G^{\mu}_\nu = -\frac14\epsilon^{\mu\rho\sigma}\epsilon_{\nu\alpha\beta} R_{\rho\sigma}^{~~\alpha\beta},
\end{equation}
along with the decomposition given in eq. \eqref{riemanndecomp}.

Projecting out the symmetric piece of eq. \eqref{dceom} precisely gives the equations of motion of topologically massive gravity
\begin{equation}\label{key}
	G_{\mu\nu} + \frac{1}{m}C_{\mu\nu} = \frac{1}{m}\epsilon^{~\alpha\beta}_{(\mu} \nabla_{\alpha} J_{\beta\nu)} + T_{\mu\nu}.
\end{equation}
Projecting out the antisymmetric piece gives
\begin{equation}\label{asymeom}
	\frac{1}{m}\epsilon_{\mu\nu\rho}\nabla_\lambda S^{\lambda\rho} = \frac{1}{4m}\epsilon_{\mu\nu\rho}\nabla^\rho R = \frac{1}{2m}\left(\epsilon^{~\alpha\beta}_\mu \nabla_{\alpha} J_{\beta\nu} - \epsilon^{~\alpha\beta}_\nu \nabla_{\alpha} J_{\beta\mu}\right).
\end{equation}
Interestingly, this equation of motion can be derived by considering a Kalb-Ramond field coupled to gravity \cite{Boldo:2002cq}, e.g. from a term in the Lagrangian like 
\begin{equation}\label{key}
	\mathcal{L}_{\text {int}} \sim \epsilon^{\mu \nu \lambda} \nabla_{\mu} B_{\nu \lambda} R \sim  \epsilon^{\mu \nu \lambda} H_{\mu\nu \lambda} R.
\end{equation}
However, since $H_{\mu\nu\rho}$ is a three-form in three-dimensions, its Hodge dual is simply a scalar, and so this ought to be equivalent to coupling a scalar to $R$. The simplest example to consider is just to say that $\epsilon^{\mu \nu \lambda} H_{\mu\nu \lambda}R = \phi R$. In \cite{Burger:2021wss,Moynihan:2020ejh}, it was shown that the propagator obtained from the double copy differs from topologically massive gravity if there is a massless mode in the spectrum. Such a propagator comes about precisely from conformally coupling a scalar to topologically massive gravity, as is shown in \cite{Deser:1990bj}.

Alternatively, we might expect something like this to arise in a dimensional reduction from string theory, where we could interpret the scalar as being a dilaton. It has shown that heterotic string theory in ten dimensions can be dimensionally reduced to $D = 3$, where it produces three-dimensional Chern-Simons terms \cite{Nishino:1991ej} and topological mass terms can indeed be produced \cite{Kaloper:1993fg}. The three-dimensional string action is given by \cite{Horowitz:1993jc}
\begin{equation}\label{key}
	S = \int d^3x \sqrt{-g} e^{-2\phi}\left[R + 4D_\mu\phi D^\mu\phi - \frac{1}{12}H_{\mu\nu\rho}H^{\mu\nu\rho}\right].
\end{equation}
The three-form field strength $H_{\mu\nu\rho}$ is proportional to the volume form in three dimensions, and it can be shown that an ansatz $H_{\mu\nu\rho} = Ne^{2\phi}\epsilon_{\mu\nu\rho}$ does in fact give rise to topological mass terms from a stringy reduction \cite{Kaloper:1993fg}. Since the double copy has a natural string-theoretic origin, this is a compelling reason to think that the analysis is correct. However, it should be noted that the taking $H_{\mu\nu\rho} \propto e^{2\phi}\epsilon_{\mu\nu\rho}$ as was done in \cite{Kaloper:1993fg} does not strictly speaking derive topologically massive gravity from string theory, since the resulting action has a Ricci scalar with the wrong sign, meaning the graviton must then be thought of as a ghost excitation.

Taking the Hodge dual of eq. \eqref{asymeom} gives an equation of motion for the Ricci scalar
\begin{equation}\label{key}
	\nabla^\mu R = 2\nabla^\mu J,
\end{equation}
where we have assumed a covariantly conserved source that satisfies $\nabla_\nu J^{\mu\nu} = 0$. However, taking the trace of eq. \eqref{dceom} gives $R = -2T$, which means for consistency we require
\begin{equation}\label{key}
	J = -T + K,
\end{equation}
where $K$ is some object that satisfies $\nabla^\mu K = 0$. We can think of this in terms of a composite graviton which contains the usual graviton, the B-field and the dilaton -- the so called fat graviton $H_{\mu\nu}$ \cite{Luna:2016hge}. In this case, the equation above tells us that we ought to have single source $T_{\mu\nu}$ that couples to both the symmetric and antisymmetric components of $H_{\mu\nu}$, e.g. we might expect an interaction term of the form
\begin{equation}\label{key}
	\cl{L}_{\text{int}} \sim \frac12 H^{(\mu\nu)}T_{\mu\nu} - \frac{1}{2m}\epsilon^{~\alpha\beta}_\mu \nabla_{\alpha} T_{\beta\nu}H^{[\mu\nu]}.
\end{equation}

Finally, we note that we can linearize eq. \eqref{dceom} around Minkowski space (considering the free case for simplicity) to find an equation of motion
\begin{equation}\label{key}
	\partial^{2} h_{\mu \nu}-\frac{1}{2} \eta_{\mu \nu} \partial^{2} h+\frac{1}{2m} \partial_{\rho} \epsilon^{\rho \lambda}_{(\mu} \partial^{2} h_{\nu) \lambda}  - \frac{1}{2m}\epsilon_{\mu\nu\rho}\pd^\rho\pd^2 h= 0.
\end{equation}
Let's now move on to investigating the duality in second-order form, this time taking \textit{all} colour indices to spacetime indices (in either the bi-adjoint scalar or in topologically massive Yang-Mills), finding
\begin{equation}\label{doubledceom}
	D^{2} \phi^{a \bar{a}}+\frac{1}{2} f^{a b c} f^{\bar{a} \bar{b} \bar{c}} \phi^{b \bar{b}} \phi^{c\bar{c}} + m^2\phi^{a\bar{a}}=J^{a \bar{a}} ~~~\longrightarrow~~~ \nabla^{2} R^{\mu\bar{\mu}} + \epsilon^{\mu\nu\rho}\epsilon^{\bar{\mu}\bar{\nu}\bar{\rho}} R_{\nu\bar{\nu}}R_{\rho\bar{\rho}}  + m^2R^{\mu\bar{\mu}} =J^{\mu\bar{\mu}}.
\end{equation}
This mapping deserves further justification. On the left-hand side, we are considering a doubly-gauged bi-adjoint scalar (gauged with respect to both barred and unbarred colours), while on the right we have the usual gravitational covariant derivative. It is useful to see precisely how these covariant derivatives are related, and so we will again make the basic colour-kinematic replacement $a\rightarrow \mu,~ \bar{a}\rightarrow\bar{\mu}$ etc to find that the covariant derivative is mapped via
\begin{equation}\label{key}
	D_\alpha\phi^{a\bar{a}} = (\pd_\alpha\phi^{a\bar{a}} + f^{abc}A_\alpha^b\phi^{c\bar{a}} + f^{\bar{a}\bar{b}\bar{c}}A_\alpha^{\bar{b}}\phi^{a\bar{c}})~~\rightarrow~~ \pd_\alpha R^{\mu\bar{\mu}} + f^{\mu\nu\rho}A_{\alpha\nu} R^{\bar{\mu}}_\rho + f^{\bar{\mu}\nu\rho}A_{\alpha\nu} R^{\mu}_\rho.
\end{equation}
We need to identify the kinematic structure constant as well as the object $A_{\alpha\nu}$. As in the gauge theory case, the first is simply the three-dimensional Lorentz algebra structure constant, i.e. the Levi-Civita tensor. The gauge field ought to be related to the spin-connection (as it was in the first-order case), and the simplest case is to identify $A_{\mu\nu}$ with the \textit{dual} spin connection\footnote{This is very similar to the four dimensions the dual spin-connection  used to derive e.g. Taub-NUT solutions (see \cite{Emond:2020lwi} eq. 3.25). There the dual spin connection is given by $\tilde{\omega}_{\mu\nu\rho} = \frac12\omega_{\mu}^{~\tau\sigma}\epsilon_{\tau\sigma\nu\rho}$.}, i.e.
\begin{equation}\label{key}
	A_{\mu\nu} = \frac12\omega_{\mu}^{~\sigma\chi}\epsilon_{\sigma\chi\nu}.
\end{equation}
Plugging this in then gives
\begin{equation}\label{key}
	D_\alpha\phi^{a\bar{a}} ~~\rightarrow~~ \pd_\alpha R^{\mu\bar{\mu}} + \omega_\alpha^{~\mu\rho}R_{\rho}^{\bar{\mu}} + \omega_\alpha^{~\bar{\mu}\rho}R_{\rho}^{\mu} = \nabla_\alpha R^{\mu\bar{\mu}}.
\end{equation}
We see then that the covariant derivatives are simply double copied under CK duality.
 
Expanding out the Levi-Civitas in eq. \eqref{doubledceom}, we find then that the double copy gives
\begin{equation}\label{key}
	(D^{2} + m^2)R^{\mu\bar{\mu}} + \left(g^{\mu\bar{\mu}}R^{\rho\sigma}R_{\rho\sigma} - 2R^{\mu}_{~\rho}R^{\rho\bar{\mu}}\right) = J^{\mu\bar{\mu}}.
\end{equation}
This is \textit{very} similar to the second-order equations of motion that we derived from topologically massive gravity, however the factor multiplying the last term is different. This has an important effect, namely that the higher-order in $R_{\mu\nu}$ terms don't vanish when we take the trace. Taking the trace of the free-field equations, we find
\begin{equation}\label{key}
	R^{\rho\sigma}R_{\rho\sigma} = -(D^2 + m^2)R.
\end{equation}

We can plug this back in to the equation of motion to find that there is an extra dynamical linear contribution
\begin{equation}\label{key}
	(D^{2} + m^2)\left(R^{\mu\bar{\mu}} - g^{\mu\bar{\mu}}R\right) - 2R^{\mu}_{~\rho}R^{\rho\bar{\mu}} = 0.
\end{equation}
Linearizing this around flat space, again in the harmonic gauge, we find
\begin{equation}\label{key}
	-(\pd^2 + m^2)\pd^2\left(h_{\mu\nu} - \eta_{\mu\nu}h\right) = 0.
\end{equation}
We see then that the double copy theory differs by a trace term, at least at linear order. This has a number of implications. Firstly, we expect the scattering amplitudes in topologically massive gravity to be equivalent to the double copy if we are scattering sources with trace-free stress-energy tensors. This means we would expect the pure topologically massive graviton amplitudes (i.e. demanding that $h_{~\mu}^\mu = 0$) to be related to pure topologically massive gluon amplitudes by BCJ duality, as was shown at three, four and five points in \cite{Gonzalez:2021bes}. We also then expect that double-copied scattering amplitudes with $T \neq 0$ to differ from those in TMG, which is in agreement with the results found in \cite{Burger:2021wss}. This is also consistent with the propagator analysis in \cite{Moynihan:2020ejh}, where it was found that the propagator double copy to topologically massive gravity only holds when the graviton is to be taken on-shell and massive with definite mass, ignoring the possibility of a $q^2 = 0$ pole. If both poles are considered to be equally plausible (as they should be for a theory with a consistent massless limit), then it was found that the residues changed and the scattering amplitudes would be markedly different, explicitly in the contribution from trace terms.

The graviton is trivially related to the linearized Ricci tensor in the harmonic gauge\footnote{While we could follow the last section and use the axial gauge for this purpose, it is more complicated in gravity and not particularly illuminating. To quote R. Delbourgo in \cite{Delbourgo:1981dt}, ``\textit{Axial gauge gravity is so complicated as to be practically useless}".}
\begin{equation}\label{key}
	 h_{\mu\nu} = \frac{2}{\pd^2}R_{\mu\nu}^L,
\end{equation} 
and thus we conclude that one-point functions are related in momentum space via
\begin{equation}\label{key}
	\braket{h_{\mu\nu}(q)}\bigg|_{J=0} = \frac{2}{m^2}\braket{R_{\mu\nu}(q)}\bigg|_{J=0}.
\end{equation}
This is strikingly similar to the topologically massive Yang-Mills case and is suggestive that we are on the right track. As a first order check of this relationship, we can proceed as we did in the last section and compute the three-particle amplitude. The vertex is given by
\begin{equation}\label{key}
	V^{\mu\bar{\mu},\nu\bar{\nu},\rho\bar{\rho}} = i\kappa m^2 \epsilon^{\mu\nu\rho}\epsilon^{\bar{\mu}\bar{\nu}\bar{\rho}},
\end{equation}
and taking $R_{\mu\nu}(q) = \frac12 m^2\tilde{\epsilon}_\mu(q)\tilde{\epsilon}_\nu(q)e^{iq\cdot x}$ we find an amplitude of the form
\begin{equation}\label{key}
	\cl{A}_3^{R^3} = \kappa\frac{m^8}{6}\left(\varepsilon^{\mu\nu\rho}\epsilon^1_\mu\epsilon_\nu^2\epsilon_\rho^3\right)^2 = \frac{m^6}{6}\cl{A}_3^{h^3}.
\end{equation}
This again matches the little-group scaling analysis and Feynman diagram approach, again in a much simpler way. Having evaluated the three-particle amplitudes, it is natural to consider more particles. However, the goal of this work is to examine the duality at the level of the equations of motion, and so this will be looked at elsewhere. One interesting point is that in standard Yang-Mills, the Feynman rules for the covariantly double copied version were not nicer than the standard Feynman rules. In 2+1 dimensions, however, this seems not to be the case, with the Feynman rules in the double copied theory appearing to be \textit{much} simpler than in the first-order formalism.
\subsection{NLSM, SG and Topologically Massive Born-Infeld Theory}
So far, we have identified the dual structure constants as belonging to the Lorentz algebra, however this isn't the only choice. As was shown in \cite{Cheung:2021zvb}, we can also choose to make a replacement dictated by the diffeomorphism algebra
\begin{equation}\label{key}
	[V\cdot\pd,W\cdot \pd] = (\cl{V}^\nu\pd_\nu \cl{W}_\mu - \cl{W}^\nu\pd_\nu \cl{V}_\mu)\pd^\mu.
\end{equation}
This motivates making the following replacement 
\begin{equation}\label{key}
	f^{abc}\cl{V}^b\cl{W}^c \rightarrow \cl{V}^\nu\pd_\nu \cl{W}_\mu - \cl{W}^\nu\pd_\nu \cl{V}_\mu.
\end{equation}
We can apply this to the ungauged massive bi-adjoint equations of motion to find a massive variant of the non-linear sigma model (NLSM), i.e. we replace $\phi^{a\bar{a}} \rightarrow j_\mu^a$ in
\begin{equation}\label{key}
	\pd^{2} \phi^{a \bar{a}}+\frac{1}{2} f^{a b c} f^{\bar{a} \bar{b} \bar{c}} \phi^{b \bar{b}} \phi^{c\bar{c}} + m^2\phi^{a\bar{a}} =J^{a \bar{a}}.
\end{equation}
This simply yields eq. \eqref{tymdc} but with $F\rightarrow j$, but without having identified the kinematic structure constant. Making the diff algebra replacement gives
\begin{equation}\label{key}
	 \pd^{2} \phi^{a \bar{a}}+\frac{1}{2} f^{a b c} f^{\bar{a} \bar{b} \bar{c}} \phi^{b \bar{b}} \phi^{c\bar{c}} + m^2\phi^{a\bar{a}} =J^{a \bar{a}} ~~~\rightarrow~~~ (\pd^{2} + m^2)j^{a}_\mu+f^{a b c} j^{b\nu}\pd_\nu j^c_\mu = \pd_\mu J^a.
\end{equation}
This is simply a massive generalisation of the NLSM found in \cite{Cheung:2021zvb}, however this mass does not appear topological in nature, and is simply related to the addition of a mass term of the form $\sim m^2j^{a\mu}j^a_\mu$ in the action. 

Making a further replacement and again using the diffeomorphism algebra gives a massive variant of the special Galileon
\begin{equation}\label{key}
	(\pd^{2} + m^2)j^{a}_\mu+f^{a b c} j^{b\nu}\pd_\nu j^c_\mu = \pd_\mu J^a ~~~\rightarrow~~~ (\pd^{2} + m^2) j_{\mu \bar{\mu}}+j^{\nu \bar{\nu}} \partial_{\nu} \partial_{\bar{\nu}} j_{\mu \bar{\mu}}-\partial^{\nu} j_{\mu \bar{\nu}} \partial^{\bar{\nu}} j_{\nu \bar{\mu}}=\partial_{\mu} \partial_{\bar{\mu}} J.
\end{equation}
Like the NLSM, the special Galileon has a standard (non-topological) mass term and is a simple generalisation of the versions derived in \cite{Cheung:2021zvb}. What is more interesting, however, is to start with topologically massive Yang-Mills in eq. \eqref{tymeom} and send colour to kinematics via $a \rightarrow \bar{\mu}$ to find
\begin{equation}\label{key}
	\partial^{\mu} F_{\mu \nu \bar{\mu}}+A^{\mu \bar{\nu}} \partial_{\bar{\nu}} F_{\mu \nu \bar{\mu}}-\partial_{\bar{\nu}} A^{\mu}{ }_{\mu} F_{\mu \nu}{ }^{\bar{\nu}} + \frac{m}{2}\epsilon_{\nu\rho\sigma}F_{\bar{\mu}}^{\rho\sigma} = \partial_{\bar{\mu}} J_{\nu}.
\end{equation}
In the $m\rightarrow 0$ limit, this is the reformulation of Born-Infeld theory derived in \cite{Cheung:2021zvb}, so it is reasonable to conclude that for $m\neq 0$ this is a reformulation of topologically massive Born-Infeld theory \cite{Tripathy:2000kv,Gaete:2003xh}. This is derived by taking the gauged bi-adjoint scalar, replacing all colour indices with kinematic and identifying the two structure constants as belonging to \textit{different} algebras: one diffeomorphism and one Lorentz.  

It would certainly be interesting to study the scattering amplitudes given by this reformulated theory, however to this authors knowledge the scattering amplitudes for the standard formulation of topologically massive Born-Infeld theory have not been worked out. Deriving the amplitudes for both formulations is beyond the scope of this paper, and so we leave this to future work.
\section{Discussion}
In this paper we have looked at topologically massive theories through the lens of covariant colour-kinematics duality. We found that the bi-adjoint scalar field is mapped to topologically massive gauge theory under the duality when replacing an SU(N) structure constant with a Lorentz group kinematic structure constant. Making the same choice for the second structure constant, we found that the gauge theory maps to topologically massive gravity coupled to a scalar or, equivalently, an anti-symmetric field. Choosing instead to consider the structure constant associated to diffeomorphisms, we found that the bi-adjoint scalar is simply related to a massive variant of the non-linear sigma model and the massive special Galileon, however these appear with non-topological mass terms. If we choose to combine two kinematic structure different constants -- one from the diffeomorphisms and one from the Lorentz group -- we land on a topologically massive generalisation of the Born-Infeld theory. This is particularly interesting, however this theory does not appear to be well studied in the literature, and it would certainly be fruitful to derive its scattering amplitudes and show that they do indeed double copy. There are several interesting follow up directions to this work. One would be to explore the supersymmetric generalisation of topologically massive gauge theories \cite{Zupnik:1988ry,1983PhLB..131...69A,Sasaki:1998rn} and the double copy to (presumably) topologically massive supergravity. This could also be interesting to look at for topologically massive Born-Infeld \cite{Bergshoeff:2014ida}. Another interesting idea might be to explore whether or not the perturbiner expansion of topologically massive gauge theories (including gravity) is related to some kind of $L_\infty$ algebra, as is the case for massless Yang-Mills, see for example \cite{Arvanitakis:2019ald,Lopez-Arcos:2019hvg}. Finally, it would be interesting to see if one can construct a set of differential operators which would transmute the tree-level amplitudes in the various different topologiclly massive theories, as was recently considered in \cite{Cheung:2017ems}. We leave these interesting ideas for the future.
\section*{Acknowledgements}
NM would like to thank William Emond, Laura Johnson, Donal O'Connell, Jann Zosso and especially Clifford Cheung for very useful discussions on this topic. NM is supported by STFC grant ST/P0000630/1 and the Royal Society of Edinburgh Saltire Early Career Fellowship. Many calculations were done using xAct \cite{xAct}.
\bibliographystyle{JHEP}
\bibliography{refs} 
\end{document}